\documentclass{article}
\usepackage{spconf,amsmath,graphicx}
\usepackage{tabularx}
\usepackage{amsmath}
\usepackage{multirow}

\title{END-TO-END SPEAKER HEIGHT AND AGE ESTIMATION USING ATTENTION MECHANISM WITH LSTM-RNN}
\name{Manav Kaushik$^{\star}$ \qquad Van Tung Pham$^{\dagger}$ \qquad Eng Siong Chng$^{\dagger}$}
\address{$^{\dagger}$ School of Computer Engineering, Nanyang Technological University (NTU), Singapore \\
      $^{\star}$Dept. of Electrical and Electronics Engineering, Birla Institute of Technology and Science (BITS) Pilani}
\begin{document}
%\ninept
%
\maketitle
%ABSTRACT:
\begin{abstract}
Automatic height and age estimation of speakers using acoustic features is widely used for the purpose of human-computer interaction, forensics, etc. In this work, we propose a novel approach of using attention mechanism to build an end-to-end architecture for height and age estimation. The attention mechanism is combined with Long Short-Term Memory (LSTM) encoder which is able to capture long-term dependencies in the input acoustic features. We modify the conventionally used Attention- which calculates context vector as the sum of attention only across timeframes- by introducing a modified context vector which takes into account total attention across encoder units as well, giving us a new cross-attention mechanism. Apart from this, we also investigate multi-task learning approach for jointly estimating speaker height and age. 
\par We train and test our model on the TIMIT corpus. Our model outperforms several approaches in the literature. We achieve a root mean square error (RMSE) of 6.92cm and 6.34cm for male and female heights respectively and RMSE of 7.85years and 8.75years for male and females ages respectively. By tracking the attention weights allocated to different phones, we find that Vowel phones are most important while Stop phones are least important for the estimation task.

\end{abstract}

\begin{keywords}
Automatic height and age estimation,  multi-task learning, Attention, Long Short-Term Memory (LSTM), Recurrent Neural Network (RNN).
\end{keywords}

%SECTION 1: Introduction
\section{Introduction}
\label{sec:intro}

Speech is a unique physiological signal which not only contains information about the linguistic content (such as words, accent, language, etc.) but also conveys the para-linguistic content (such as height, age, gender, emotions, etc.). This helps us in estimating the physical parameters like
height and age of a speaker, which holds a wide variety of applications in the real-world such as natural human-machine interaction, speaker profiling, and forensics \cite{tanner2004forensic, articleSchuller}.

A typical approach for speaker characteristic estimation is to apply shallow learning techniques, such as linear regression \cite{Williams2013SpeakerHE} or support vector machine \cite{poorjam2015height, mahmoodi2011age, bocklet2008age}, on top of utterance-level representation such as i-vector \cite{poorjam2015height, bocklet2008age} or x-vector \cite{ghahremani2018end}. Such approaches are not end-to-end since the utterance-level representation extractors are trained separately for speaker recognition tasks which are not optimized for height and age estimation. \par In this work, we propose an end-to-end approach for speaker height and age estimation. The acoustic features are encoded using an LSTM network which allows capturing long-term dependencies. The uniqueness and novelty of our technique come from two aspects. Firstly, we propose to use attention mechanism for speaker characteristic estimation task. As from our best knowledge, there is not any work in the literature which studies the use of attention for speaker height and age estimation. Secondly, instead of performing attention across speech frames, as done conventionally \cite{xu2015show, shan2018attention}, we perform attention across both speech frames and encoder units to obtain two context vectors then combine them to generate a final context vector. We believe that the proposed attention, denoted as cross attention, captures much more information than the conventional counterpart and hence could produce better performance. 

\par By analyzing attention weights across speech frames, we find that that highest weights have been assigned to Vowel phones while the lowest weights have been assigned to Stop phones. Since speaker characteristics such as height and age are correlated with the length of speaker vocal tract \cite{singh2016short, smith2005interaction}, this higher attention to vowel phones maybe attributed to the fact that these phones require the use of the glottal vocal tract while utterance of stop phones does not involve its use. 
\par Apart from attention, we study how passing in gender information as a feature helps the model to better estimate the height and age of a speaker.

%SECTION 1: Literature Review
\section{RELATED WORKS}
\label{sec:lit_review}

Most of previous studies on height and age estimation tend to use conventional approaches of applying shallow learning techniques. For instance, Williams et al. \cite{Williams2013SpeakerHE} combine Gaussian Mixture Models (GMM) and linear regression subsystems to estimate the speaker height. Poorjam et al. \cite{poorjam2015height} and Bahari et al. \cite{bahari2012age} predict speaker height and age by applying least-squares Support Vector Regression (SVR) on top of i-vector. Mahmoodi et al. \cite{mahmoodi2011age} use Support Vector Machines (SVMs) while Bocklet et al. \cite{bocklet2008age} use GMM supervectors with SVM for age estimation task. Singh et al. \cite{singh2016short} use a bag of words representation generated from short-term cepstral features and train a Random Forest regressor for age and height estimation. The issue with the above mentioned approaches is that none of them are end-to-end modeling techniques and thus, are not specifically optimized to speaker physical parameter estimation such has height and age.
\par More recently, Ghahremani et al. \cite{ghahremani2018end} propose an end-to-end deep neural network (DNN) for age prediction while Kalluri et al.\cite{kalluri2019deep} also attempt to jointly predict both height and age of speaker using a unified end-to-end DNN model which is initialized using a conventional system based on SVR trained with Gaussian Mixture Model-Universal Background Model (GMM-UBM) supervector features. Although, both of these works employ an end-to-end architecture of their estimation tasks, they rely more on conventional approaches. 
\par Attention mechanism\cite{bahdanau2014neural} has been successfully applied for various research topics such as neural machine translation, keyword spotting \cite{shan2018attention}, and computer vision \cite{mnih2014recurrent, ba2014multiple, xu2015show}. To our best knowledge, none of the past works in the literature have used attention for speaker  physical parameter estimation. Our work is the first in the literature to demonstrate the potential of attention mechanism in tracking the relational dependency and importance of different phones in estimating speaker height and age in an utterance.
% SECTION 3: Dataset
\section{DATASET USED}
\label{sec:dataset}

We use the TIMIT dataset \cite{garofolo1993timit} for all our experiments done in this study. TIMIT has a total of 6300 unique utterances. There are 630 speakers of these utterances who are distributed across 8 different dialect regions with each speaker speaking ten different utterances. The gender distribution of the speakers in male to female is 2:1. Moreover, the dataset also includes time-aligned orthographic, phonetic and word transcriptions which help us track phonetic attention.
\par The train-test split is given in the dataset i.e. 461 speakers (326 male and 135 female) for training and validation, and 162 speakers (112 male and 56 female) for testing. The height of speakers in the training data ranges from 145cm to 199cm and in testing data, they range from 153cm to 204cm. Similarly, the age of speakers ranges from 21 years to 76 years in training data and 22 years to 68 years in test data. There is no overlapping of speakers between test and training datasets. Moreover, the duration of the utterances ranges from 1- 6s with an average of about 2.5s.

% SECTION 4: Methods
\section{METHODOLOGIES}
\label{sec:methodologies}

Our proposed framework for end-to-end speaker height and age estimation is shown in Fig. 1. First, data preprocessing is applied on the input audio to generate acoustic features such as Filter bank energies and pitch. Then these features are encoded by an LSTM network before feeding them to an attention layer. Subsequently, the output of attention layer, which is a vector, is transformed by a dense layer to make height and age prediction. In following subsections, we describe these processes in detail.

\subsection{Data Preprocessing and Augmentation}
\label{ssec:data_pre}

Our input for each utterance is a two-dimensional matrix consisting of $T$ timeframes with each timeframe consisting of 83 acoustic features (80 filter bank and 3 pitch features) extracted from windows of 25ms with 10ms stride. We apply Cepstral Mean and Variance Normalization (CMVN) to these acoustic features. The resulting features are ${\bf x}=[x_1,x_2,…,x_T ]$   where each frame, $x_i$, contains  83 features.
\par Since neural network models require a huge amount of data to train properly, we use speed perturbation as an augmentation step to obtain the audio signals at 1.1x and 0.9x speeds as well. Apart from this, we use spectral augmentation (SpecAugment) \cite{park2019specaugment} to enhance the robustness of our model by randomly masking strands from feature and time axes covering approximately 10-12\% of the training data.

\begin{figure*}[ht]
\centering
\includegraphics[width = 0.7 \textwidth]{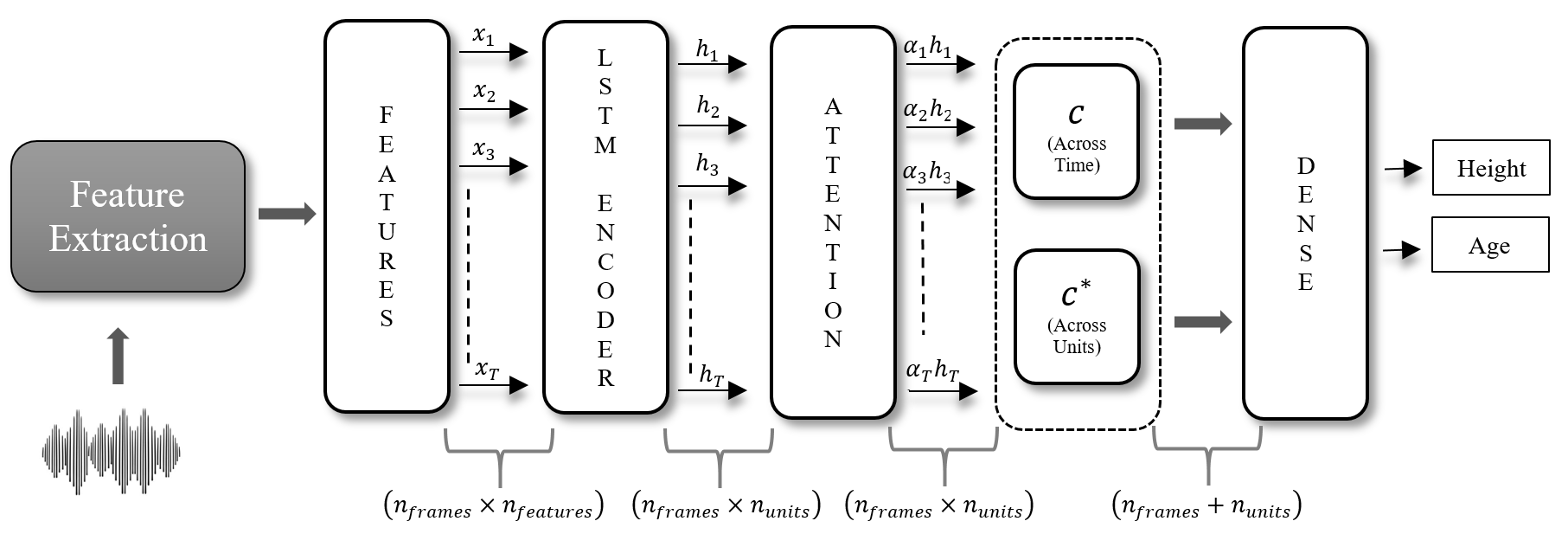}
\caption{The proposed framework for speaker height and age estimation}
\end{figure*}

% SUBSECTION 4.1: LSTM
\subsection{LSTM RNN Encoding}
\label{ssec:lstm_encoding}
Since LSTM has shown to be efficient to capture long temporal dependencies, we chose this architecture to encode acoustic features for our study. Given a sequence of input features ${\bf x}=[x_1,x_2,…,x_T ]$, the LSTM network processes it frame-by-frame to generate the sequence of hidden states $ {\bf h}=[h_1,h_2,….,h_T ]$ where each state $h_i$ has dimension of $n_{units}$ i.e. number of LSTM units. 
\par Once the input features are encoded, the straightforward approach is to take the final hidden state i.e. $h_t$ as the utterance-level representation for height and age estimation. We denote this setting as "last hidden state" for the subsequent experiment in Table 1 in Section \ref{ssec:attention}. However, in practice, LSTM tends to forget information when operated on longer sequences. Therefore, we propose to use attention mechanism to solve this problem as presented in Section \ref{ssec:attention}.

% $n_{timeframes}$ × $n_{features\; per\; frame}$
% $n_{timeframes}$ × $n_{lstm\; units}$

% SUBSECTION 4.2: Attention
\subsection{Attention Mechanism}
\label{ssec:attention}

The attention mechanism was primarily introduced to help memorizing long sentences in neural machine translation \cite{bahdanau2014neural}. It generates a context vector as weighted sum of hidden states of the LSTM encoder over all timeframes. Since the attention mechanism has access to the entire input sequence, the problem of forgetting initial parts of the sequence is solved. We use soft attention which is typically used in previous works \cite{xu2015show, shan2018attention}.

First, a scalar score $e_t$ is estimated for each LSTM hidden state $h_t$  as:
\begin{equation}
e_t = {\bf v_a}^{\tau}tanh({\bf W_a}h_t + {\bf b})
\end{equation}
% \[ e_t = {\bf v_a}^{\tau}tanh({\bf W_a}h_t + {\bf b}) \]

where ${\bf v_a}$, ${\bf W_a}$ and ${\bf b}$ are learnable parameters.
Then, the attention weights $\alpha_t$ are obtained by applying a softmax function on $e_t$, i.e. 
\begin{equation}
\alpha_t = \frac{exp(e_t)}{\sum_{i=1}^{T} exp(e_i)} 
\end{equation}

Since we use a softmax function, $\alpha_t  \epsilon [0,1]$ and \( {\sum_{t=1}^{T}\alpha_t} = 1\).
After this, we obtain a context vector, $\bf c$, as the weighted average across all timeframes of the LSTM outputs $\bf h$:
\begin{equation}
{\bf c} = \sum_{t=1}^{T} \alpha_t h_t
\end{equation}

As a result, $\bf c$ has the same dimension as hidden states $h_t$ i.e. $n_{units}$
\par Instead of considering $\bf c$ as the final context vector (which is the conventional approach \cite{xu2015show, shan2018attention}), we propose a cross-attention approach in which we further perform attention across all the $n_{units}$ LSTM units to generate another context vector, denoted as $\bf c^*$, and concatenate them to obtain the final context vector $\bf f$.

\begin{equation}
{\bf f} = [{\bf c}, {\bf c^*}]
\end{equation}

Note that $\bf c^*$ has dimension of $n_{frames}$, hence $\bf f$ has dimension of $(n_{frames}$ + $n_{units})$. $\bf f$ is finally passed into a dense layer which makes the final prediction.

% SUBSECTION 4.3: Dense
\subsection{Dense Layers}
\label{ssec:dense}

For each of height or age, the estimation made as:
\begin{equation}
{\hat y} = ReLU( {\bf v^\tau}{\bf f})
\end{equation}

where $\bf v$ is a learnable vector of length same as $\bf f$. We use Mean Squared Error (MSE) as our loss function:
\begin{equation}
\textstyle loss = \textstyle \frac{1}{N}[\sum_{i=1}^{N}{(y_i - \hat{y_i})^2}]
\end{equation}

where $y_i$ and $\hat {y_i}$  are the actual and predicted values respectively for utterance $i$ and $N$ is the total number of utterances.

We also study multi-task learning which aims to estimate height and age at the same time. The training loss for our approach is calculated as:
\begin{equation}
 Loss_{total} = a(loss_{height}) + (1-a)(loss_{age} )
 \end{equation}
where $a$ is a hyper-parameter optimized on the validation set.
% \par We further study how to leverage gender information to improve our models. We study 3 approaches:
% \begin {itemize}
%     \item Training two separate attention models for male and female. To overcome the problem of limited gender-dependent data, we initialize both the models with weights pre-trained using the entire training data and then fine-tune these models separately using gender-specific data. We name this gender-dependent model as $GD_{separate}$
%     \item Treating gender information as a single binary feature and concatenating it with the acoustic features. We name it as $GD_{featConcat}$
%     \item Adding gender classification as one more task. We name this gender-independednt model as $GI_{genPrediction}$ . In this case, our training loss, $L$ is: 
%     \[ L = a(loss_{height}) + (b)(loss_{age}) + (1-a-b)(loss_{gender} ) \]
% \end {itemize}

% SECTION 5: RESULTS
\section{EXPERIMENTAL RESULTS}
\label{sec:res}

% SUBSECTION 5.1: Experimental Setup
\subsection{Experimental Setup}
\label{ssec:setup}

The LSTM network consists of 1 hidden layer with 128 units. Moreover, we use a dropout regularization of 20\% and a recurrent dropout of 20\% to avoid overfitting. We also apply dropout regularization of 20\% on the dense layer.

\par For the performance analysis of models, we use standard metrics of Root Mean Squared Error (RMSE) and Mean Absolute Error (MAE), which are defined as:
\begin{equation}
RMSE = \sqrt{\frac{\sum_{i=1}^{N} (y_i - \hat{y_i})^2}{N}} \;;\; MAE = \frac{\sum_{i=1}^{N}|y_i - \hat{y_i}|}{N}
\end{equation}

where $y_i$ and $\hat {y_i}$ are the actual and predicted values respectively of $i$-th utterance and $N$ is the total number of utterances. 

% \par We compare our different experiment with our baseline and work by Singh et al. [18] with 25ms analysis window.

% SUBSECTION 5.2: Quantitative Results
\subsection{Quantitative Results}
\label{ssec:results}

In the subsequent experiments, we analyze the performance
of different models from the literature and different variation
of attention mechanism that we propose (at an analysis window of 25ms). 
% As shown in tabele 1, most of the works in literature use different models for male and female data (as denoted as 'Separate Models' in Table 1) while we train a single model for the complete dataset.

% We first compare results of different attention techniques: (1)last hidden state (mentioned in Section \ref{ssec:lstm_encoding}); (2)Conventional attention \cite{xu2015show, shan2018attention}; (3) Proposed cross attention. We use multi-task learning for all three attention techniques. We then compare results of multi-task with that of single-task. Finally, we show the effect of combining gender information (binary 0/1 value) with acoustic features. Results are shown in Table 1. We observe that the proposed cross attention outperforms other techniques in all conditions. We also see multi-task learning tends to enhance the generalization ability of the model and thus, gives better results compared to single-task. Finally, adding gender information consistently improves performance of our model. 

% We then take our best model, i.e. multi-task cross attention with gender information, and compare it to other models from the literature (considering analysis window of 25ms for all models). Our results outperforms several approaches in literature and is competitive with the state-of-the-art as shown in Table 2. The advantage of our approach is to use a single end-to-end model for both height and age estimation.

% Please add the following required packages to your document preamble:
% \usepackage{multirow}

\begin{table*}[h]
\centering
\begin{tabular}{cccccccc}
% \centering
\hline\hline
\multirow{2}{*}{\bf \emph{Techniques}}               & \multirow{2}{*}{\bf \emph{Multi-task?}} & \multirow{2}{*}{\bf \emph{Gender Consideration}} & \multirow{2}{*}{} & \multicolumn{2}{c}{\bf \emph{Height (cm)}} & \multicolumn{2}{c}{\bf \emph{Age (years)}} \\ %\cline{5-8} 
                                         &                              &                                      &                   & \emph{RMSE}           & \emph{MAE}         & \emph{RMSE}        & \emph{MAE}        \\ \hline\hline
\multirow{2}{*}{\emph {Last hidden state}}       & \multirow{2}{*}{Yes}         & \multirow{2}{*}{Not considered}                  & Male              & 8.23              &  6.86           &  9.24           &    7.03        \\ %\cline{4-8} 
                                         &                              &                                      & Female            &   7.94            & 6.19            &   10.20          &  7.71          \\ \hline
\multirow{2}{*}{\emph{Conventional-Attention}} & \multirow{2}{*}{Yes}         & \multirow{2}{*}{Not considered}                  & Male              &  6.99             &  5.42           &  8.08           & 5.84           \\ %\cline{4-8} 
                                         &                              &                                      & Female            &  6.62             & 5.36            &   9.08          &   6.24         \\ \hline
\multirow{2}{*}{\emph{Cross-Attention}}         & \multirow{2}{*}{No}         & \multirow{2}{*}{Not considered}                  & Male              & 6.98              &  5.38           & 8.16            & 5.97           \\ %\cline{4-8} 
                                         &                              &                                      & Female            &  6.56             &   5.30          &   9.12          &    6.27        \\ \hline
\multirow{2}{*}{\emph{Cross-Attention}}         & \multirow{2}{*}{Yes}          & \multirow{2}{*}{Not considered}                  & Male              &  6.94             &  5.29           &  7.90           &  5.62          \\ %\cline{4-8} 
                                         &                              &                                      & Female            &   6.40            & 5.15            &   8.87          & 6.16           \\ \hline
\multirow{2}{*}{\bf \emph{Cross-Attention}}         & \multirow{2}{*}{\bf Yes}         & \multirow{2}{*}{\bf As a binary feature}                  & \bf Male              &  \bf 6.92             &  \bf 5.24           &  \bf 7.85           & \bf 5.62           \\ %\cline{4-8} 
                                         &                              &                                      & \bf Female            &    \bf 6.34           & \bf 5.09            & \bf 8.75            & \bf 6.08           \\ \hline

\multirow{2}{*}{\emph{Singh et al. \cite{singh2016short}}}         & \multirow{2}{*}{No}         & \multirow{2}{*}{Separate Models}                  & Male              &  6.9             &  5.2           &  8.0           &  5.7          \\ %\cline{4-8} 
                                         &                              &                                      & Female            &   6.3            &  5.1           &  8.8           &  6.1         \\ \hline

\multirow{2}{*}{\emph{Kalluri et al. \cite{kalluri2019deep}}}         & \multirow{2}{*}{Yes}         & \multirow{2}{*}{Separate Models}                  & Male              & 6.85              &  -           &  7.60           & -           \\ %\cline{4-8} 
                                         &                              &                                      & Female            &  6.29             &  -           &   8.63          & -           \\ \hline
                                         
\multirow{2}{*}{\emph{Mporas et al. \cite{mporas2009estimation}}}         & \multirow{2}{*}{No}         & \multirow{2}{*}{Separate Models}                  & Male              &  6.8             &  5.3           &  -           &  -          \\ %\cline{4-8} 
                                         &                              &                                      & Female            &   6.3            &  5.1           &  -           &  -         \\ \hline

\multirow{2}{*}{\emph{Williams et al. \cite{Williams2013SpeakerHE}}}         & \multirow{2}{*}{No}         & \multirow{2}{*}{Separate Models}                  & Male              & -              &  5.37           &  -           & -           \\ %\cline{4-8} 
                                         &                              &                                      & Female            &  -             &  5.49           &   -          & -           \\ \hline\hline
\end{tabular}
\caption{Comparison of the end-to-end framework with different settings and literature} % title name of the table
\label{tab:PPer}
\end{table*}

% \par From Table 1 we may deduce that Multi-task learning tends to enhance
% the generalization ability of the model and thus, gives
% better results compared to single-task model. It can also
% be seen that the proposed cross-attention significantly outperforms
% conventional attention \cite{xu2015show, shan2018attention}. Moreover, our results outperform several approaches in literature and is competitive with the state-of-the-art as shown in Table 1. The advantage of our approach is to use a single end-to-end model for both height and age estimation.
We first compare results of different attention techniques: last hidden state (mentioned in Section \ref{ssec:lstm_encoding}); Conventional attention \cite{xu2015show, shan2018attention}; proposed cross attention with single-task and multi-task learning.
Then we show the effect of combining gender information (binary 0/1 value) with acoustic features. Results are shown in Table 1. 
\par We observe that the proposed cross attention outperforms other techniques in all conditions. We also see multi-task learning tends to enhance the generalization ability of the model and thus, gives better results compared to single-task. Finally, adding gender information consistently improves performance of our model. Comparing with other works from the literature, it may be seen that our proposed model outperforms several approaches in literature and is competitive with the state-of-the-art. The advantage of our approach is to use a single end-to-end model for both height and age estimation.

% SUBSECTION 5.3: Analysis
\subsection{Analysis}
\label{ssec:analysis}

\par We study which phones are important for age and height estimation by tracking the attention weights for each phone in the TIMIT data. We note that TIMIT contains manual phone boundaries, therefore, we can infer phone labels for each timeframe of an utterance. We accumulate the weight across all utterances to obtain an average weight for each phone.  
\par There are a total of 60 different phones which have been used in the TIMIT dataset, and a broader analysis of attention weight distribution shows us that highest attention weights have been assigned to Vowel phones (such as ay, aw, aa, ae, ao, eh, ey, etc. as shown in Fig. 2) which actually hold the highest amount of linguistic and para-linguistic information as they involve the use of glottal vocal tract \cite{singh2016short} while lowest weights have been assigned to Stop phones (such as d, b, p, k ,etc.) and some of the Fricatives (such as s, sh, f, etc.) which do not involve use of glottal vocal tract.

\begin{figure}[ht]
% \centering
\includegraphics[width = 0.52 \textwidth]{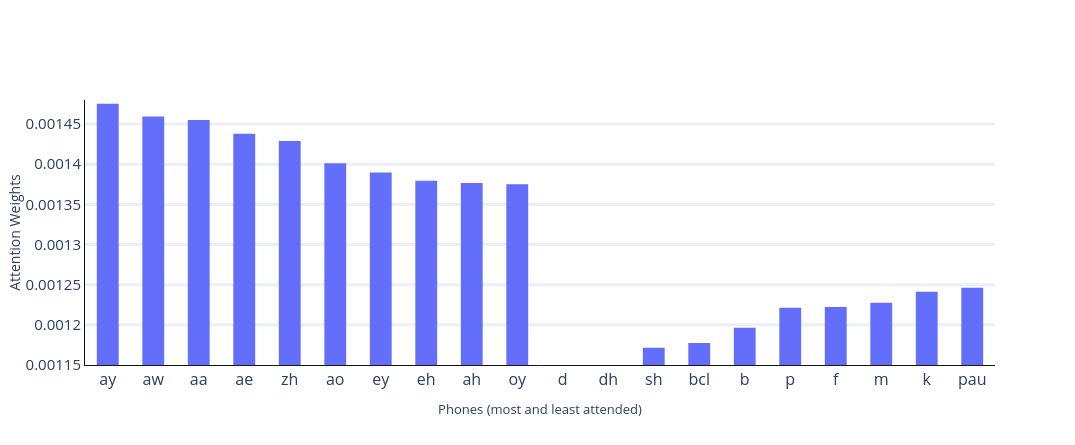}
\caption{10 Highest \& 10 Least Attended Phones respectively}
\end{figure}

% SECTION 6: Conclusions
\section{CONCLUSIONS}
\label{sec:conclusions}

We have proposed a cross-attention approach for the task of joint speaker height and age estimation. The proposed approach not only performed attention across timeframes but also performed attention across hidden units which produces more informative context vector. Experimental results on TIMIT data shows that our proposed approach combined with multi-task learning outperforms many models in the literature. Furthermore, by tracking attention weights across timeframes, we found that Vowel phones are most important while Stop phones have been least attended by the soft attention model for speaker physical characteristics estimation.

\newpage
\newpage

% SECTION 7: References
% \section{REFERENCES}
% \label{sec:refs}

% List and number all bibliographical references at the end of the
% paper. The references can be numbered in alphabetic order or in
% order of appearance in the document. When referring to them in
% the text, type the corresponding reference number in square
% brackets as shown at the end of this sentence \cite{chorowski2015attention}. An
% additional final page (the fifth page, in most cases) is
% allowed, but must contain only references to the prior
% literature.

% References should be produced using the bibtex program from suitable
% BiBTeX files (here: strings, refs, manuals). The IEEEbib.bst bibliography
% style file from IEEE produces unsorted bibliography list.
% -------------------------------------------------------------------------
\bibliographystyle{IEEEbib}
\bibliography{strings}

\end{document}